\documentclass[useAMS,usenatbib]{mn2e}
\usepackage{natbib}
\usepackage{color,graphicx} 
\usepackage{amsmath}        
\usepackage{amsfonts}       
\usepackage{amssymb}        
\usepackage{wasysym}        

\begin{document}
\author[K.~Liu et al.]
{K.~Liu,$^{1,2}$ G.~Desvignes,$^{4}$ I.~Cognard,$^{2,1}$
B.~W.~Stappers,$^{3}$ J.~P.~W.~Verbiest,$^{5,4}$
\newauthor K.~J.~Lee,$^{6,4}$ D.~J.~Champion,$^{4}$ M.~Kramer,$^{4,3}$
P.~C.~C.~Freire,$^4$ and R.~Karuppusamy,$^4$\\
  $^{1}$Station de radioastronomie de Nan\c{c}ay, Observatoire de
  Paris, CNRS/INSU, F-18330 Nan\c{c}ay, France \\
  $^{2}$Laboratoire de Physique et Chimie de l'Environnement et de l'Espace
  LPC2E CNRS-Universit\'{e} d'Orl\'{e}ans, F-45071 Orl\'{e}ans Cedex
  02, France \\
  $^{3}$University of Manchester, Jodrell Bank Centre of Astrophysics,
  Alan-Turing Building, Manchester M13 9PL, UK\\
  $^{4}$Max-Planck-Institut f\"{u}r Radioastronomie, Auf dem H\"{u}gel
  69, D-53121 Bonn, Germany \\
  $^{5}$Fakult\"{a}t f\"{u}r Physik, Universit\"{a}t Bielefeld, Postfach 100131, 33501 Bielefeld, Germany\\
  $^{6}$KIAA, Peking University, Beijing 100871, P.R. China \\}
\title[2-D template-matching]{Measuring pulse times of arrival from broadband pulsar observations}

\maketitle

\begin{abstract}
In recent years, instrumentation enabling pulsar observations with
unprecedentedly high fractional bandwidth has been under development
which can be used to substantially improve the precision of pulsar
timing experiments. The traditional template-matching method used to
calculate pulse times-of-arrival (ToAs), may not function
effectively on these broadband data due to a variety of effects such
as diffractive scintillation in the interstellar medium, profile
variation as a function of frequency, dispersion measure (DM)
evolution and so forth. In this paper, we describe the channelised
Discrete Fourier Transform method that can greatly mitigate the
influence of the aforementioned effects when measuring ToAs from
broadband timing data. The method is tested on simulated data, and
its potential in improving timing precision is shown. We further
apply the method to PSR~J1909$-$3744 data collected at the
Nan\c{c}ay Radio Telescope with the Nan\c{c}ay Ultimate Pulsar
Processing Instrument. We demonstrate a removal of systematics due
to the scintillation effect as well as improvement on ToA
measurement uncertainties. Our method also determines temporal
variations in dispersion measure, which are consistent with
multi-channel timing approaches used earlier.
\end{abstract}


\section{Introduction}\label{sec:intro}
High-precision pulsar timing is the essential tool in the currently
ongoing gravity tests with binary pulsars
\citep[e.g.][]{fwe+12,afw+13}, and gravitational wave detection with
pulsar timing arrays \citep[e.g.][]{hd83,fb90}. Developments in
observing backends based on polyphase filter bank techniques now
allow pulsar observations with large fractional bandwidth
\citep[e.g. $\gtrsim1/3$, see][]{rdf+09,sha+11,ctg+13}. Broadband
systems such as the Ultra-broadband (UBB) receiver developed at the
Effelsberg radio telescope which achieves an instantaneous frequency
coverage from 600\,MHz to 3\,GHz are being implemented \citep{k+14}.
Extending the observing bandwidth greatly increases the
signal-to-noise ratio (S/N) of the observed pulsar profiles which
can be translated into a decrease in measurement uncertainties of
pulse times-of-arrival \citep[ToAs,][]{dr83}. This is likely to
improve the resulting pulsar timing precision, under the
circumstance that the actual timing residuals are dominated by white
noise \citep{cs10,lvk+11}.

Typically, determination of ToAs is achieved by first correcting for
any frequency-dependent delays caused by dispersion in the
interstellar medium, then averaging the data across all frequency
channels and finally cross-correlating the resulting pulse profile
with a template \citep{tay92}. The template can either be a high-S/N
profile obtained directly from observations or an analytic model of
the pulse \citep[e.g.][]{kxc+99}. However, applying this approach to
broadband pulsar timing data may be problematic for a number of
reasons as follows:
\begin{itemize}
\item The pulse profile shape is not constant across the observing
bandwidth. This can be due to intrinsic profile variation across
frequency, frequency-dependent scattering timescales, instrumental
channelisation and so forth. When the observed flux density at
different frequencies varies in time due to interstellar
scintillation, the shape of the frequency-averaged profile will be
changed depending on which part of the band scintillates up and
which part scintillates down. In turn, this variation in profile
shape will affect the cross-correlation and therefore the derived
ToAs \citep[e.g.][]{lvk+11}.
\item In many cases the pulsar
dispersion measure\footnote{Dispersion measure is defined as the
integrated electron density between the Earth and the pulsar.} (DM)
has been witnessed to evolve in time \citep[e.g.][]{kcs+13}. Hence,
applying a constant but not perfectly correct DM value for
dedispersion would smear the profile shape differently as the DM
values vary, and thus shift the ToAs. Using an incorrect DM can also
induce misalignment of profiles from different frequencies, which
would mimic profile variation across the observing band and enhance
the scintillation effect mentioned above.
\item Variations in the radio frequency interference (RFI) environment can change the usable
frequency channels in different observations. This effectively
changes the frequency range from which the frequency-averaged
profile is composed of due to the frequency-dependence of the pulse
profile shape. Similar variations in the frequencies used (and
therefore in the resulting profile shape) could be caused by
instrumental failure or slight changes to the observing system.
\end{itemize}
Consequently, an alternative approach that utilises the data's
frequency information to avoid the corrupting effects listed above,
is required. Without frequency-averaging, any frequency-dependent
changes in brightness will no longer result in changes in the
overall pulse shape. DM variations can also be taken into account
effectively if the data cover a large enough frequency range.

Besides the current standard template-matching scheme in
\cite{tay92}, there are also alternative methods targeting different
issues. \cite{hbo05a} applied Gaussian interpolation when
cross-correlating profiles with low S/Ns. \cite{van06} used full
Stokes information rather than only the total intensity for ToA
determination. \cite{ovh+11} decomposed profiles in principal
components to improve ToA measurements when the timing residuals are
dominated by pulse phase jitter. In this paper we develop a method
to measure ToAs based on timing data with frequency information.

The structure of this paper is as follows. In
Section~\ref{sec:algrim} we describe our method to generate ToAs
with frequency-resolved data. In Section~\ref{sec:simu} we present
results from tests based on simulated data. Application of the
method to real data is shown in Section~\ref{sec:real}. We conclude
with a brief discussion in Section~\ref{sec:conclu}.

\section{Methodology} \label{sec:algrim}
In the traditional approach, the observed pulse profile $P(t)$, is
described by a one dimensional array modelled by
\begin{equation}
P(t)=a+bS(t-\Delta\tau)+n(t), \label{eq:mod_t}
\end{equation}
where $S(t)$ is the standard template obtained from previous
observations, $a$ is the baseline difference, $b$ is a scaling
factor, $\Delta\tau$ is the phase difference between the profile and
the template, and $n(t)$ is a noise component. The actual fitting
routine is often carried out in the Fourier domain, where after a
discrete Fourier transform (DFT), the model can be written as
\citep{tay92}
\begin{equation}
P_ke^{i\theta_k}=aN\delta_{k}+bS_ke^{i(\phi_k+k\Delta\tau)}+G_k,~~~k=0,...,N-1,
\label{eq:mod_f}
\end{equation}
where $P_k$ and $S_k$ are the amplitudes of the complex Fourier
coefficients, $\theta_k$ and $\phi_k$ are the phases, $\delta_{k}=1$
($k=0$) or 0 ($k\neq0$), $N$ is the number of frequency bins, and
$G_k$ represents random noise equal to the Fourier transform of the
sampled noise in the time-domain profile, $n(t)$. The best estimated
value of $\Delta\tau$, can then be found by minimising the
goodness-of-fit statistic
\begin{equation}
\chi^{2}(b,\tau)=\sum^{N-1}_{k=1}\left|\frac{P_{k}-bS_{k}e^{i(\phi_{k}-\theta_{k}+k\Delta\tau)}}{\sigma_{k}}\right|^{2},
\end{equation}
where $\sigma_{k}$ is the root-mean-square intensity of the noise at
frequency $k$.

If the frequency resolution is kept, the data arrays have to be
expanded into a second dimension and in this case the dispersive
delay of different frequency bands also needs to be taken into
account. Mostly, the delay is found to fulfill the scaling of
$t_{\rm d}\propto{\rm DM}/f^2$, where $f$ is the observing frequency
\citep{lk05}\footnote{Further propagation effects in the
interstellar medium \citep[e.g. scattering, as shown in][]{bcc+04}
could introduce other scalings, too, albeit at a less significant
level. A quantitative summary of such is not within the scope of
this paper, but can be found in e.g. \cite{lk05}.}. Accordingly, one
can express the two-dimensional model as:
\begin{equation}
P(f,t)=a(f)+b(f)S(f,t-\Delta\tau-\mathcal{D}\times\Delta{\rm
DM}/f^2)+n(f,t), \label{eq:mod_t}
\end{equation}
where $\mathcal{D}$ is the dispersion constant and $\Delta{\rm DM}$
is the difference in dispersion measure between the profile and the
template. \footnote{In this paper we use the definition of
$\mathcal{D}\equiv1/K$ where $K\equiv2.410\times10^{-4}$\,$\rm
MHz^{-2}~cm^3~pc~s^{-1}$.} Note that $\mathcal{D}\times\Delta{\rm
DM}/f^2$ vanishes when $f\rightarrow\infty$. Hence, $\Delta\tau$ can
be directly related to the ToA at infinite frequency, which is
obtained after correcting the dispersion delay.

Similar to the treatment in the one-dimensional case, one can carry
out a DFT individually on each frequency band, and perform a
two-dimensional fit over the entire bandwidth for $\Delta\tau$ and
$\Delta{\rm DM}$. The approach will be later referred to as the
``channelised DFT method''. After the transformations, the model in
the frequency domain can be written as
\begin{equation}
P_{j,k}e^{i\theta_{j,k}}=a_jN\delta_{k}+b_jS_{j,k}e^{\phi_{j,k}+k\tau_j}+G_{j,k},
\end{equation}
where $k=0,...,N-1$ and $j=1,...,N_{\rm b}$; $N$ is the number of
bins, $N_{\rm b}$ is the number of frequency bands, and
\begin{equation}
\tau_j=\frac{\mathcal {D}\times\Delta
DM}{f^2_j}\frac{2\pi}{P}+\Delta\tau.
\end{equation}
Here $\tau_j$ is in units of radians and $P$ is the pulsar's
rotational period. The data of different frequencies are assumed to
be from a single observation, which means they are aligned in time
by default and thus $\Delta\tau$ is constant over the entire band.
Following \cite{tay92}, $a_j$ is immediately obtained from
\begin{equation}
a_j=\frac{P_{j,0}-b_jS_{j,0}}{N}.
\end{equation}
The best estimates of $\Delta\tau$ and $\Delta{\rm DM}$ can then be
obtained by minimising the function
\begin{equation}
\chi^2(b_i,\Delta\tau,\Delta{\rm DM})=\sum^{N_{\rm
b}}_{j=1}\sum^{N-1}_{k=1}\left|\frac{P_{j,k}-b_jS_{j,k}e^{i(\phi_{j,k}-\theta_{j,k}+k\tau_j)}}{\sigma_{j,k}}\right|^2.
\label{eq:chisqr_DM}
\end{equation}
Here the same as in \cite{tay92}, the likelihood estimator is
calculated using all informative frequency harmonics\footnote{In
\cite{tay92}, the calculation of $\chi^2$ includes frequency bins
only to $N/2$ because the rest are simply a symmetric duplication
and would not add any useful information. Decreasing the number of
harmonics used for the fitting may risk discarding useful
information and thus affecting measurement precision, especially
when profiles are narrow or have sharp features so that the high
frequency bins contain significant power.}. In order to perform the
fitting, we modified the Levenberg-Marquardt (L-M) routine in
\cite{ptvf92} to adopt a model of complex numbers, and recalculated
the curvature matrix in this case. More details can be found in
Appendix~\ref{sec:app}. The $1$-$\sigma$ errors of the fitted
parameters are determined from the covariance matrix which is
obtained by taking the inverse of the curvature matrix. A similar
approach based on the same likelihood estimator as in
Eq.~(\ref{eq:chisqr_DM}), using however a different fitting routine,
has been developed by \cite{pdr14}.

\begin{figure}
\centering
\includegraphics[scale=0.85]{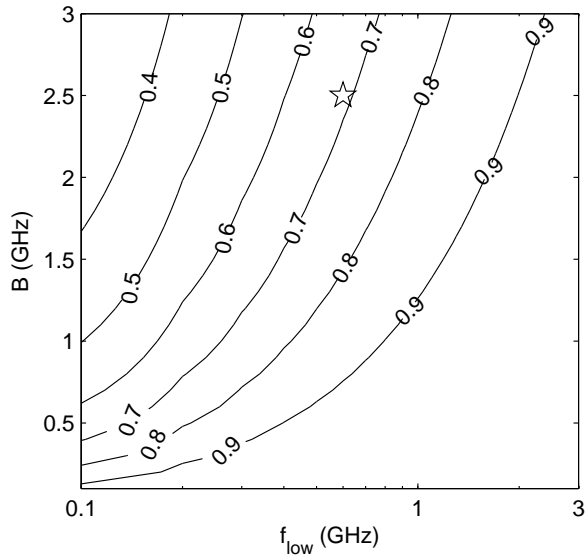}
\caption{Correlation coefficient ($\rho$) between $\Delta\tau$ and
$\Delta{\rm DM}$ for different bandwidths and frequency ranges. Here
$f_{\rm low}$ is the frequency at the lower edge of the bandwidth,
$\nu$ is the total bandwidth. The Effelsberg UBB setting is
represented by the star. \label{fig:f_bw}}
\end{figure}

It can be seen from the expression of $\chi^2$ in
Eq.~(\ref{eq:chisqr_DM}), that for data with a bandwidth that does
not result in a significant difference between $f_1$ and $f_{N_{\rm
b}}$, $\Delta\tau$ and $\Delta{\rm DM}$ would be highly correlated.
Following the calculation in \cite{lbj+14}, in Fig.~\ref{fig:f_bw}
we plot the correlation coefficient ($\rho$) between the two
parameters for a given observing bandwidth and lower bound of
observing frequency. Clearly, $\rho>0.9$ for all current L-band
(1-2\,GHz) timing observations (with bandwidth $<1$\,GHz), meaning
that solely with those data one cannot significantly break the
degeneracy between these two parameters. In this case, fitting for
both parameters would greatly worsen the accuracy of the obtained
value. The UBB receiver on the Effelsberg 100-m radio telescope can
significantly decrease the level of correlation, resulting in $\rho$
less than 0.7. Nevertheless, when the observing bandwidth is not
enough to be sensitive to $\Delta{\rm DM}$ fit, one can always
choose to fit solely for $\Delta\tau$. Under this circumstance, the
likelihood estimator is simplified into
\begin{equation}
\chi^2(b_i,\Delta\tau)=\sum^{N_{\rm
b}}_{j=1}\sum^{N-1}_{k=1}\left|\frac{P_{j,k}-b_jS_{j,k}e^{i(\phi_{j,k}-\theta_{j,k}+k\Delta\tau)}}{\sigma_{j,k}}\right|^2.
\label{eq:chisqr_noDM}
\end{equation}

\section{Simulations} \label{sec:simu}
In this section, the channelised DFT method is tested based on
simulated data. Here we carried out two types of tests: one to
assess the functionality of the algorithm in general, and the other
to evaluate the potential improvement to ToA measurements when
applied in a realistic scenario.

\subsection{Test of algorithm functionality} \label{ssec:test_func}
Ideally, a functioning template-matching approach is able to measure
the offsets in phase and DM between the profile and the template, as
well as their 1-$\sigma$ uncertainties. In the following tests, we
simulate data with a bandwidth of 500 MHz, between 1.2 and 1.7\,GHz.
A Gaussian shape with no frequency dependence is assumed for the
two-dimensional template. Individual observations are simulated by
adding white noise to the standard profile, until a given S/N is
reached. We hereby define the S/N as the ratio of the peak amplitude
of the pulse and the rms in the off-pulse region, and use this
definition throughout this paper.

\subsubsection{Consistency of parameter recovery and uncertainty} \label{sssec:test_para_rec}
\begin{figure}
\centering
\includegraphics[scale=0.5,angle=-90]{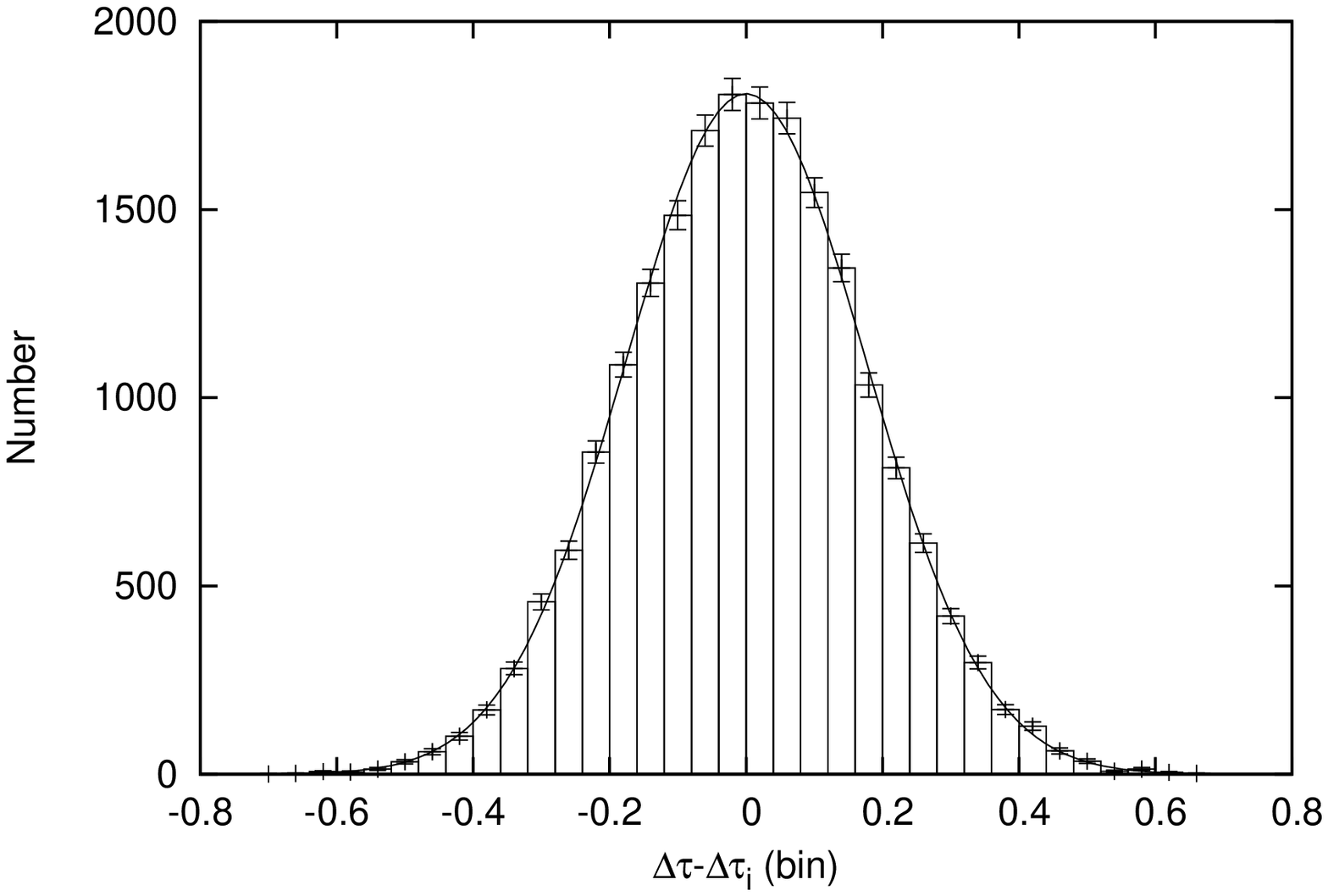}
\includegraphics[scale=0.5,angle=-90]{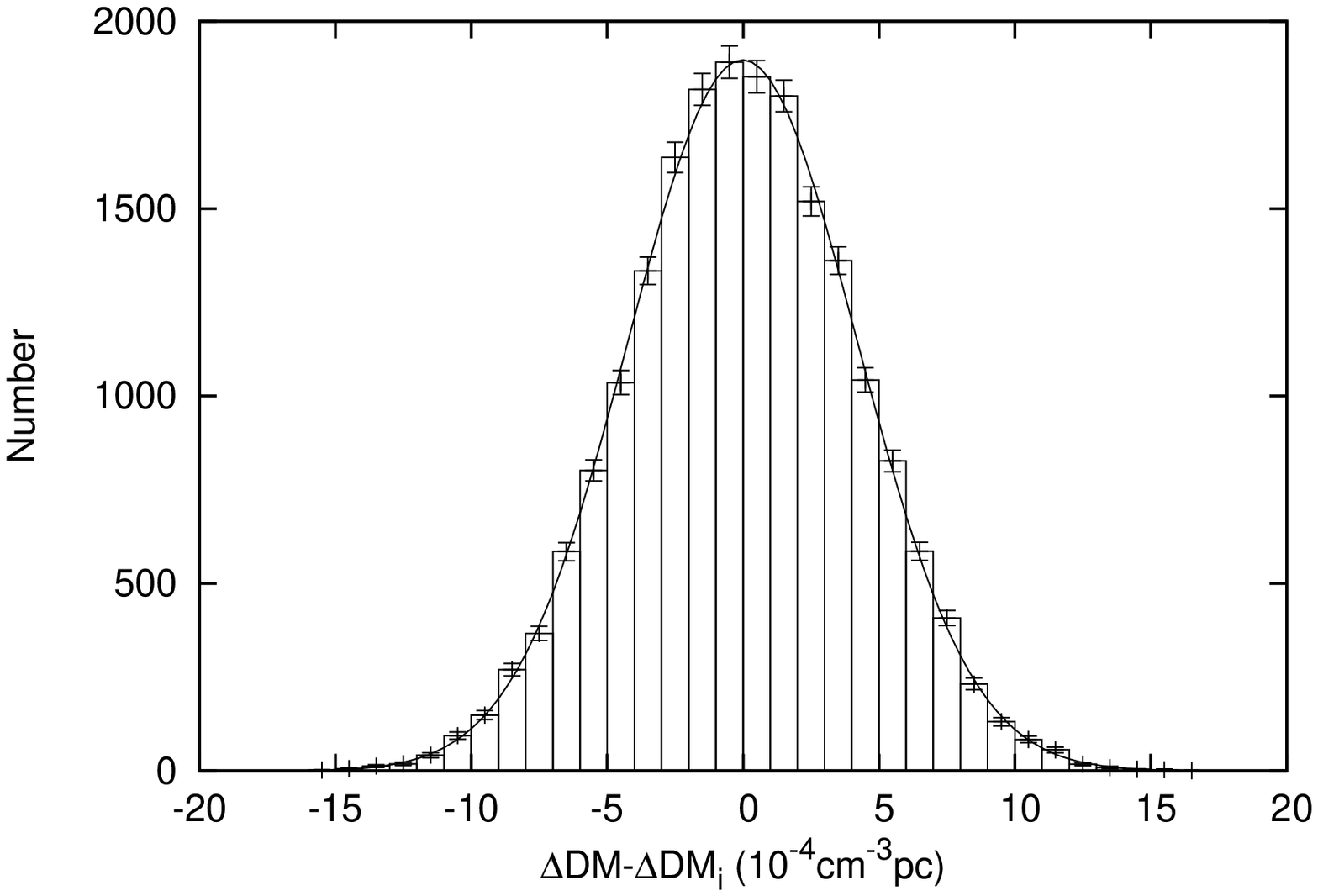}
\caption{Histograms of measured $\Delta\tau$ (top) and $\Delta{\rm
DM}$ (bottom) from $2\times10^4$ profiles after subtraction of the
input values ($\Delta\tau_i$ and $\Delta{\rm DM}_i$, separately).
The profiles were simulated with S/N=100, 16 frequency channels, and
randomly distributed $\Delta\tau$ and $\Delta{\rm DM}$ with respect
to the template. The distributions are well described by Gaussian
functions, with reduced $\chi^2$ of fit close to unity (Top: 0.96;
Bottom: 0.92). The rms in the two cases are 0.176\,bins and
$4.20\times10^{-4}$\,cm$^{-3}$pc, respectively, consistent with the
measurement uncertainties which are 0.177\,bins,
$4.22\times10^{-4}$\,cm$^{-3}$pc, respectively. \label{fig:MCtest}}
\end{figure}

In order to show that the method measures the parameters
consistently with the calculated errors, we simulated $2\times10^4$
profiles with randomly distributed offsets in phase and DM with
respect to the two-dimensional template. Fig.~\ref{fig:MCtest} shows
histograms of measured $\Delta\tau$ (top) and $\Delta{\rm DM}$
(bottom) values from the channelised DFT method, after subtraction
of the input offsets ($\Delta\tau_i$ and $\Delta{\rm DM}_i$,
separately). In both cases, the distribution is well described by a
Gaussian function and the rms is consistent with expected from the
measurement uncertainty.

\subsubsection{Consistency of measurement with theoretical
expectation} \label{sssec:test_expec}

\begin{figure}
\centering
\includegraphics[scale=0.5,angle=-90]{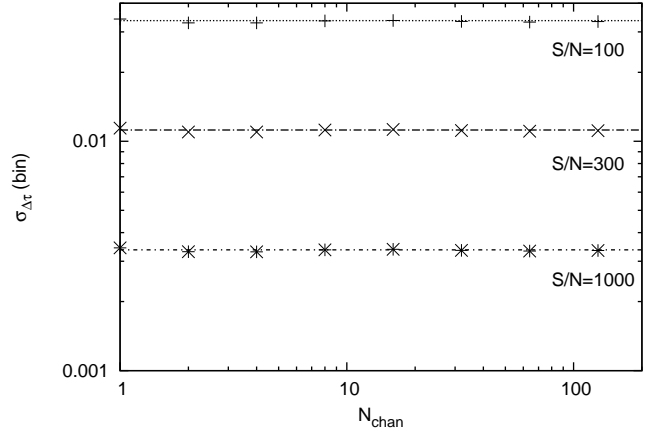}
\caption{$\sigma_{\Delta\tau}$ obtained from the channelised DFT
method for profiles of different S/N values and numbers of channels.
The lines represent the expected values from the radiometer equation
for different S/Ns. The fractional differences between the
precisions and the expectations are all within
2.5\%.\label{fig:test-fs}}
\end{figure}

\begin{figure}
\centering
\includegraphics[scale=0.5,angle=-90]{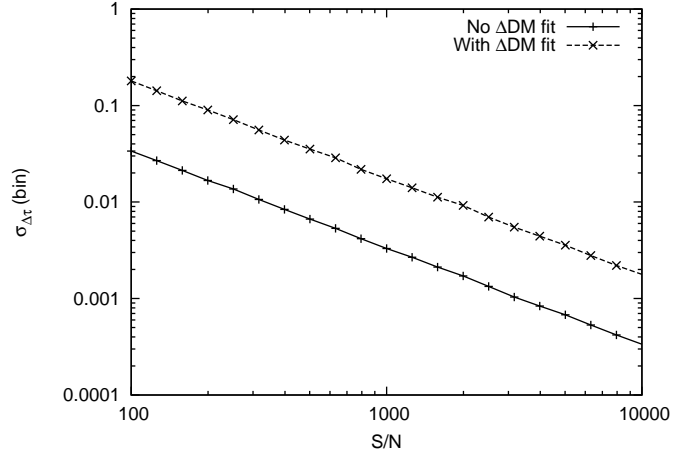}
\caption{$\sigma_{\Delta\tau}$ obtained from the channelised DFT
method for profiles with different S/Ns. Both cases are presented
where $\Delta{\rm DM}$ is simultaneously fitted or not.
\label{fig:test-snr}}
\end{figure}

The intrinsic uncertainty of a ToA is expected to be only related to
the S/N and the profile shape, and to scale inversely with the S/N
\citep{dr83}. In order to show that the method measures the phase
offset with the expected uncertainty, we simulated profiles with
different S/Ns and with different numbers of frequency channels.
Fig.~\ref{fig:test-fs} shows the measured phase uncertainties
$\sigma_{\Delta\tau}$ obtained from these profiles. The values were
compared with expected measurement uncertainties calculated from the
radiometer equation as in \citet{dr83}. Clearly, the uncertainties
for data of identical S/N all fully agree with the expectation, and
the number of channels has no effect (as is to be expected for a
pulse profile with no frequency evolution). Note that for this test
the fit of $\Delta{\rm DM}$ has been switched off, as the strong
correlation (corresponding to $\rho=0.982$ in Fig.~\ref{fig:f_bw})
between $\Delta{\rm DM}$ and $\Delta\tau$ would greatly worsen the
measurement uncertainty. In this case it is not expected that the
resulting ToA uncertainties would be the same as calculated from the
radiometer equation. Fig.~\ref{fig:test-snr} shows the scaling of
$\sigma_{\Delta\tau}$ with S/N, as well as the influence of
simultaneous $\Delta{\rm DM}$ determination. It can be seen that the
scaling follows an inverse trend in either case, with or without
simultaneous $\Delta{\rm DM}$ fitting. When both $\Delta{\rm DM}$
and $\Delta\tau$ are fitted, due to the high degree of correlation
between these two parameters the measured uncertainties increase
substantially (here by a factor of 6).

\subsubsection{Reliability of measurement uncertainty in low-S/N regime} \label{sssec:test_lowS/N}
\begin{figure}
\centering
\includegraphics[scale=0.5,angle=-90]{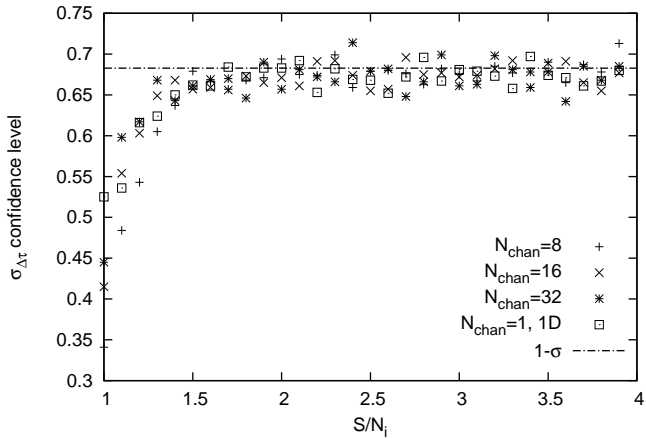}
\caption{Confidence level of $\sigma_{\Delta\tau}$ by the
channelised DFT method in the low S/N regime, based on profiles of
different numbers of frequency channels ($N_{\rm chan}$) and S/N
values in individual channels (S/N$_{i}$). The result achieved with
the traditional method (1D) is also shown for comparison.
\label{fig:MCtest-snr}}
\end{figure}

It is known that in the low S/N regime (e.g. S/N close to unity) the
traditional template-matching technique may fail to recover the true
phase offset within the measured uncertainty \citep[e.g.][]{liu12}.
Fig.~\ref{fig:MCtest-snr} demonstrates that this is also true for
the channelised DFT method. For this test, we generated pulse
profiles with the number of frequency channels, $N_{\rm chan}$ equal
to 8, 16 and 32 respectively, and varied the S/N per frequency
channel, S/N$_i$, between 1 and 4. For each combination of $N_{\rm
chan}$--S/N$_i$, we simulated $10^3$ profiles and determined how
many of these resulted in measurements consistent with the input
value for the phase offset (defined by falling into the range
decided by the measurement uncertainty), while fitting for
$\Delta\tau$ and $\Delta\rm DM$ simultaneously. This fraction is
shown on the y-axis of Fig.~\ref{fig:MCtest-snr}. Since the
uncertainties are 1-$\sigma$ values, one would expect all points in
this plot to fall around a value of 0.68. This is clearly not the
case for S/N$_i$ values less than 2.0, and the deviations from 0.68
occur at the same S/N$_i$ value for different channel numbers. The
same situation was seen when investigating the consistency of the
measured $\Delta\rm DM$ values. Statistics of $\Delta\tau$
measurements based on the traditional technique show deviation from
1-$\sigma$ at the same S/N$_i$ value. Therefore, for more reliable
measurements in this case, either frequency channels need to be
combined to increase the S/N$_i$, or an alternative approach has to
be considered \citep[as in e.g.][]{hbo05a}.

\subsubsection{Impact of template with finite S/N} \label{sssec:test_nostmpl}
\begin{figure}
\centering
\includegraphics[scale=0.5,angle=-90]{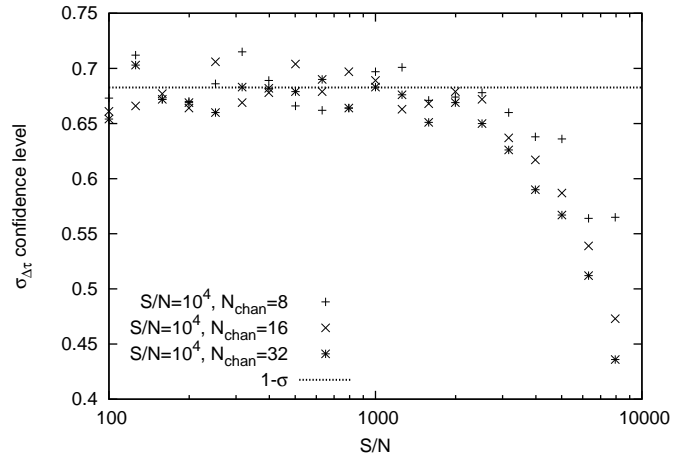}
\caption{Confidence level of $\sigma_{\Delta\tau}$ from the
channelised DFT method, based on templates with a S/N$=10^4$ and
different numbers of channels. \label{fig:MCtest-nostmpl}}
\end{figure}

As shown in \cite{lvk+11}, using a template which is not entirely
noise free may also limit the functionality of the template-matching
technique. This issue is demonstrated in
Fig.~\ref{fig:MCtest-nostmpl}, where we evaluated the accuracy of
the measured uncertainties using the same approach as the one
followed for Fig.~\ref{fig:MCtest-snr}, but this time with standard
templates of S/N=$10^4$ and different $N_{\rm chan}$ values. It can
be seen that the $\sigma_{\Delta\tau}$ values prove to be reliable,
but only until the simulated profiles reach a S/N that is within an
order of magnitude of the S/N of the template profile. Therefore,
the algorithm works as expected when the template profile has a
sufficiently high S/N and the observations have a significantly
lower S/N. Also in this case, the impact of the number of frequency
channels is limited, though fewer frequency channels (i.e. with
higher $S/N_i$ values) do make for slightly more reliable
measurement uncertainties. Again, the $\Delta\rm DM$ measurements
were seen to be influenced by the noise of the template in the same
manner.

\subsection{Expected improvement on real data}
\begin{figure}
\centering
\includegraphics[scale=0.5,angle=-90]{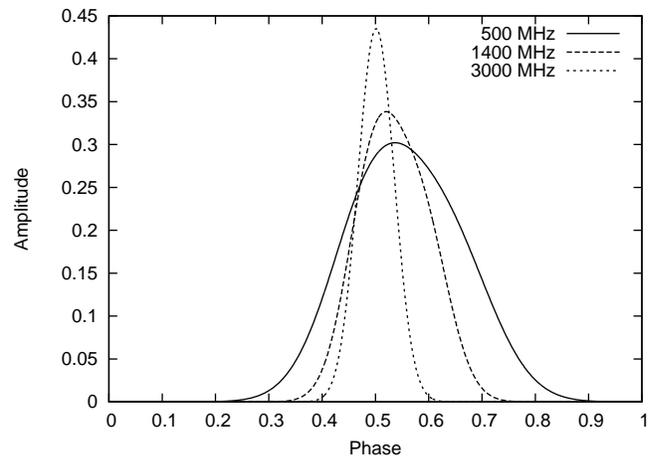}
\caption{Examples of frequency-dependent template shape at three
frequencies. Note that when later used to create profiles, the
template amplitudes from different frequencies were modulated based
on a spectral index of -1.8. \label{fig:tmpls}}
\end{figure}

As discussed in Section~\ref{sec:intro}, applying the traditional
template-matching approach to broadband pulsar timing data would not
properly produce ToAs in some circumstances. Below we demonstrate
improvements that are achieved using the channelised DFT method.
Here we simulate profiles and conduct the fitting based on a
frequency-dependent template model and examples can be found shown
in Fig.~\ref{fig:tmpls}.

\subsubsection{Accounting for DM variations} \label{sssec:test_DM}
\begin{figure}
\centering
\includegraphics[scale=0.35,angle=-90]{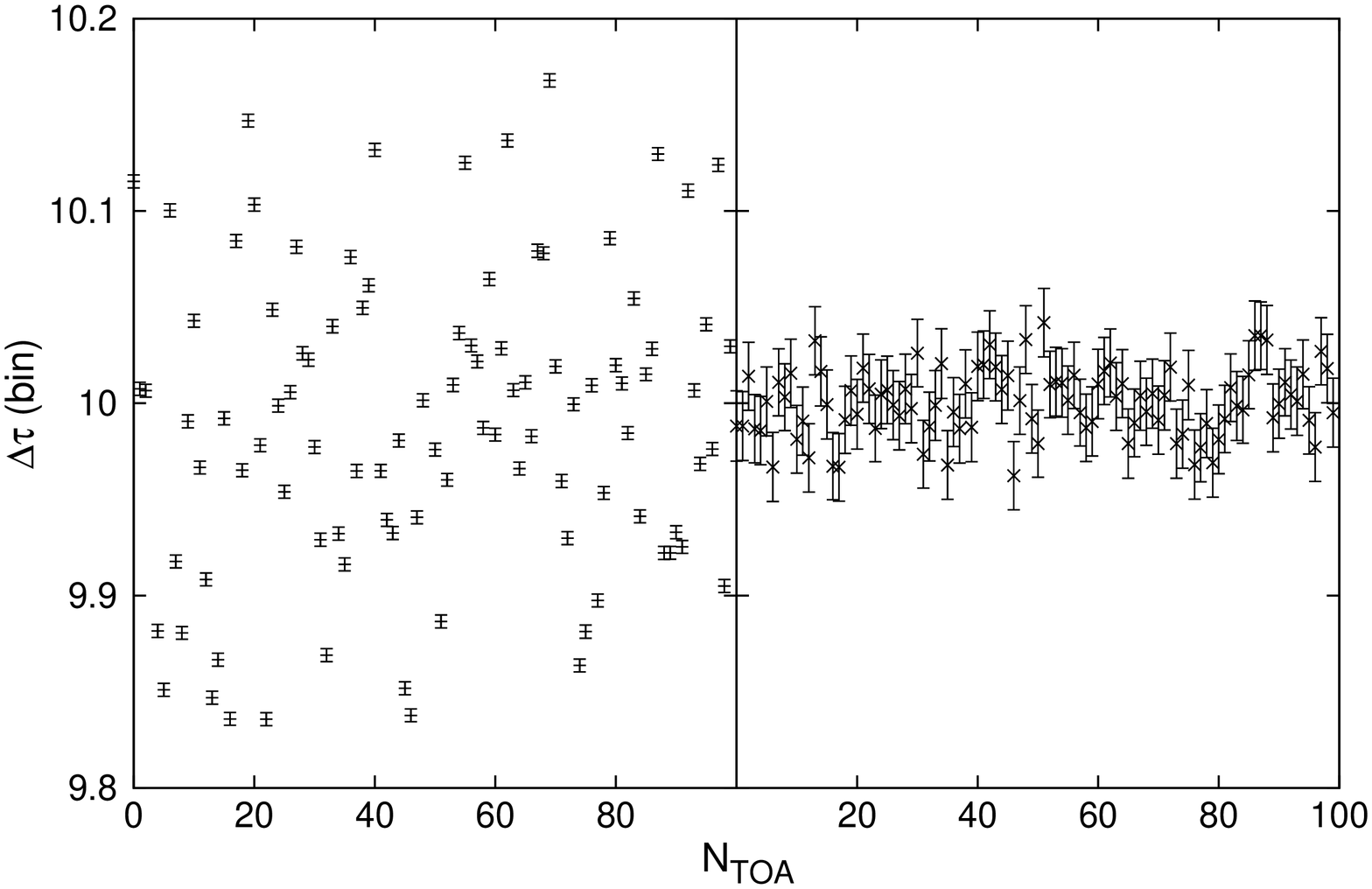}
\includegraphics[scale=0.5,angle=-90]{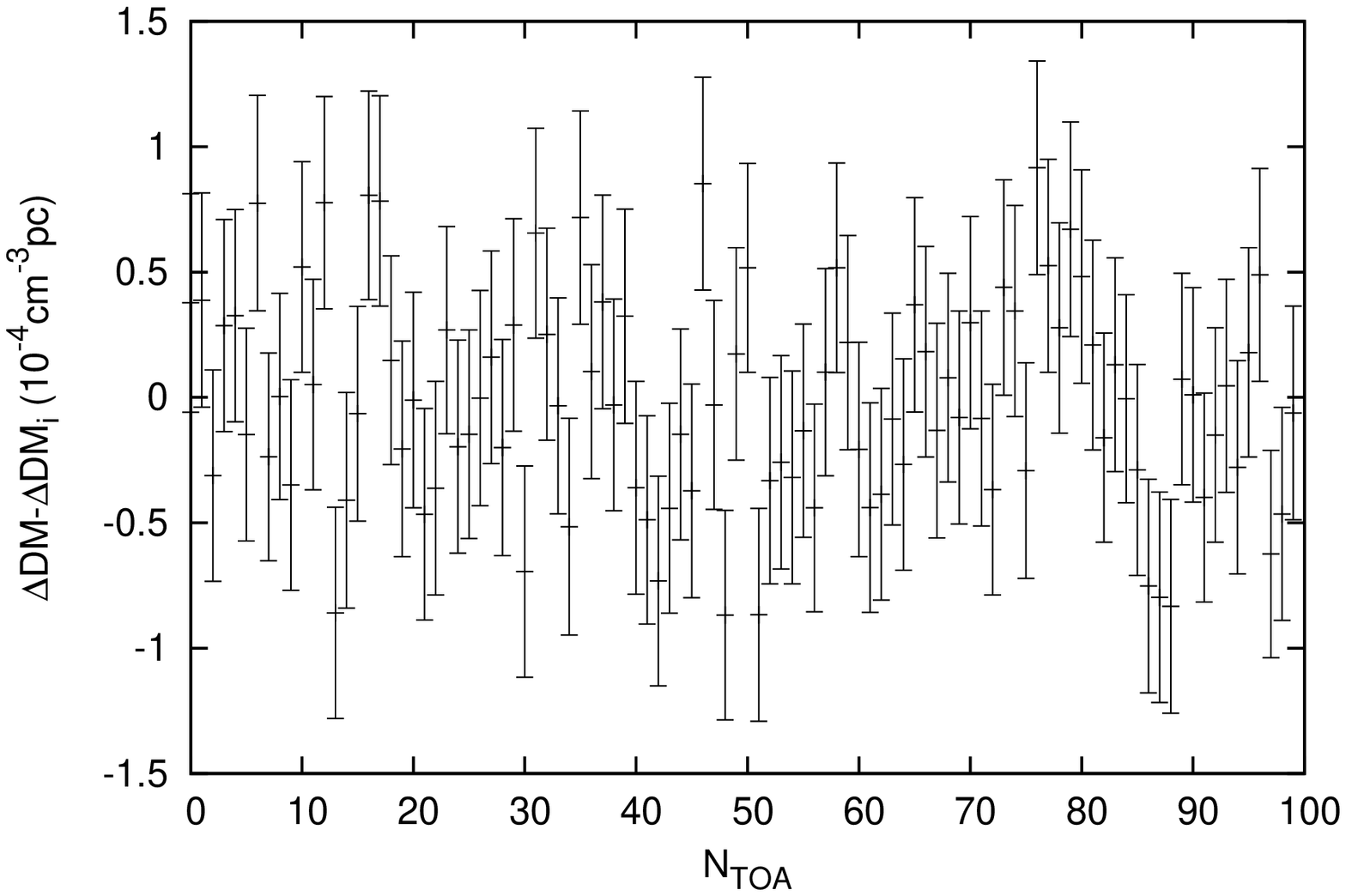}
\caption{Top:Fitted $\Delta\tau$ values of 100 profiles whose DM
offsets from the template follow a Gaussian distribution with
standard deviation $2\times10^{-4}$\,$\rm cm^{-3}pc$. The assumed
observing bandwidth is 500\,MHz, from 1.2\,GHz to 1.7\,GHz. The
input phase offset is 10\,bins. Here we used a S/N of $10^3$ and
kept 16 frequency bands when fitting for both $\Delta\tau$ and
$\Delta{\rm DM}$. The fit is carried out both without (left panel)
and with (right panel) a fit for $\Delta{\rm DM}$. The measurements
including a $\Delta{\rm DM}$ fit have a reduced $\chi^2$ of 1.01,
while those without result in a reduced $\chi^2$ of order $\sim$650.
Bottom: Fitted $\Delta{\rm DM}$ from the top right plot after
subtraction of the input DM offset, $\Delta{\rm DM}_i$. The
corresponding reduced $\chi^2$ is 1.02. \label{fig:MCtest-DMvar}}
\end{figure}

When the DM of a pulsar varies between observations, using a
constant DM value for dedispersion will introduce a bias in the
measured ToAs. As discussed in Section~\ref{sec:algrim}, when the
observing bandwidth is sufficiently large, use of a two-dimensional
template can take the influence of variable DMs into consideration
when calculating the ToAs. For demonstration purposes, we simulated
100 profiles with Gaussianly distributed DM variations\footnote{Note
that the observed DMs are more likely to show gradual variations
\citep[e.g.][]{kcs+13}. Nevertheless, our simulation is enough to
demonstrate the improvement by performing a simultaneous fit for
$\Delta\tau$ and $\Delta{\rm DM}$.} with a standard deviation of
$2\times10^{-4}$\,$\rm cm^{-3}pc$. The simulated data have a
bandwidth of 500\,MHz between 1.2 and 1.7\,GHz. The offsets in phase
were measured using the channelised DFT method, both with and
without a simultaneous fit for $\Delta{\rm DM}$. From the results in
Fig.~\ref{fig:MCtest-DMvar} (top plot, left panel), it is clear that
the obtained $\Delta\tau$ values are scattered beyond the tolerance
of the measurement uncertainties if the variable DM values are not
accounted for. On the contrary, fitting for both parameters results
in $\Delta\tau$ and $\Delta{\rm DM}$ values consistent with the
input, and with reduced $\chi^2$ values close to unity for both
measurement sets.

\subsubsection{Reducing broadband data} \label{sssec:test_bb}
\begin{figure}
\centering
\includegraphics[scale=0.42]{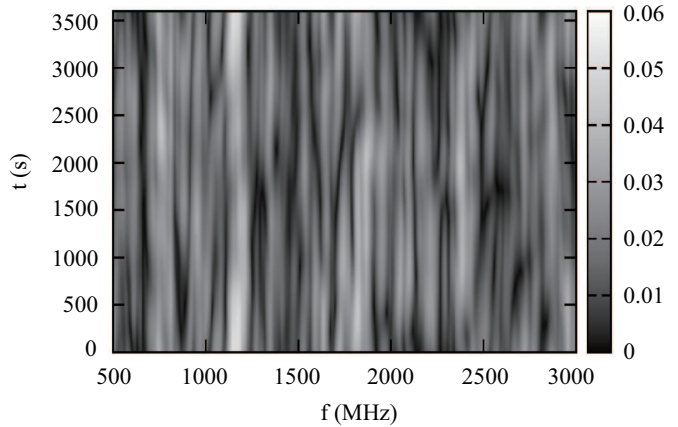}
\caption{Example simulated dynamic spectrum with length 1\,hr and
frequency coverage from 500\,MHz up to 3\,GHz. The time and
frequency resolution are 10\,s and 5\,MHz, respectively. The scale
is given on the right hand side. \label{fig:dyspec}}
\end{figure}

As mentioned in Section~\ref{sec:intro}, the profile shape
variability related to diffractive scintillation would be most
significant for broadband pulsar timing data. In order to
demonstrate the improvement in timing precision with the channelised
DFT method in this case, we simulated data covering a frequency
range from 500\,MHz up to 3\,GHz. In total we created 1-hr
observations on 100 epochs. For each epoch, to better simulate real
observations, we generated a dynamic spectrum assuming a
scintillation timescale of 20\,min and frequency scale of
50\,MHz\footnote{Note that when generating the dynamic spectra we
did not consider the dependency of the scintillation timescale and
frequency upon observing frequency. However, the simulation is
sufficient for demonstrating the potential improvement from the new
algorithm.}, respectively. An example of such a spectrum can be
found in Fig.~\ref{fig:dyspec}. To create data at a given epoch, we
firstly generated profiles for every 10\,s and 5\,MHz, based on the
template shape in Fig.~\ref{fig:tmpls}. Here we assumed a profile
S/N of 20 at 1.4\,GHz based on 5\,MHz bandwidth and a 1-hr
integration. Then the amplitudes of the profiles were weighted
differently with respect to the dynamic spectrum. Next the profiles
were integrated over the whole 1-hr session to be used for ToA
measurements. When performing the channelised DFT method we kept a
frequency resolution of 50\,MHz.

\begin{figure}
\centering
\includegraphics[scale=0.35,angle=-90]{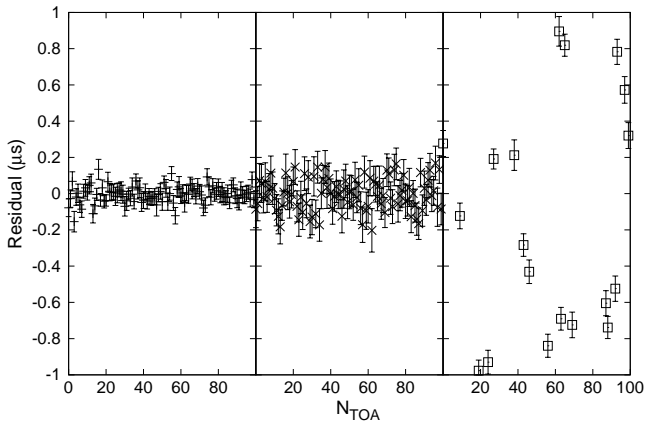}
\caption{Measurements of $\Delta\tau$ from 100 simulated profiles,
which include a shape dependency with observing frequency and flux
variation due to diffractive scintillation. Applying the channelised
DFT method without a $\Delta{\rm DM}$ fit (left panel) leads to
measurement residuals of RMS 47\,ns and reduced $\chi^2$ of 1.07.
Including a simultaneous $\Delta{\rm DM}$ fit (middle panel) reduces
the precision to 90\,ns while the corresponding measurement reduced
$\chi^2$ remains close to unity (0.97). The application of the
traditional template-matching approach to the data after averaging
over all frequency channels (right panel) results in a RMS of
4.3\,$\mu$s and a reduced $\chi^2$ of approximately 600. Not all of
the data points are therefore shown in the plot.
\label{fig:fulltest}}
\end{figure}

The results are summarised in Fig.~\ref{fig:fulltest}. It can be
seen that using the channelised DFT method in both modes (with and
without $\Delta{\rm DM}$ fit) succeeds in obtaining measurement
residuals with reduced $\chi^2$ close to unity. The factor of nearly
2 difference in the measurement precision is due to the correlation
between $\Delta\tau$ and $\Delta{\rm DM}$ which still corresponds to
a correlation coefficient of approximately 0.65. Simply applying the
traditional template-matching approach to the data after averaging
over all frequency channels, leads to precision more than two orders
of magnitude worse and a reduced $\chi^2$ value of approximately
600.

\section{Application to real data} \label{sec:real}
\subsection{Data} \label{ssec:obs}
Timing observations of millisecond pulsars have been conducted
regularly at the Nan\c{c}ay Radio Telescope (NRT). The legacy
Berkeley-Orleans-Nan\c{c}ay (BON) backend has been in operation for
nearly ten years, capable of producing 128\,MHz data which are
coherently dedispersed online \citep{ct06}. The Nan\c{c}ay Ultimate
Pulsar Processing Instrument (NUPPI), which started in late 2011, is
a baseband recording system using a Reconfigurable Open Architecture
Computing Hardware (ROACH) FPGA board developed by the CASPER
group\footnote{http://casper.berkeley.edu/} and GPUs. Here the
analog-to-digital converters firstly sample and digitise the signal
over a 512\,MHz band at the Nyquist rate with dual polarisations in
8-bit. Then a polyphase filter bank is performed to channelise the
band into 128 channels each of 4\,MHz width. Next the channels are
packetised into four sub-bands and sent to four GPU clusters
individually via 10\,GbE links for online coherent dedispersion, at
a DM of 10.3940\,$\rm cm^{-3}pc$ for PSR~J1909$-$3744. Finally, the
data were folded to form 1-min integrations.

NUPPI data of PSR~J1909$-$3744 provide a good opportunity to test
the application of the channelised DFT approach, as this pulsar is
known to show significant flux variation due to interstellar
scintillation with frequency scale of order $\sim50$\,MHz
\citep[e.g.][]{cl02}, and thus less than the observing bandwidth.
Accordingly, we chose data collected from $\sim30$ epochs between
MJD~56545 and 56592, with central frequency at 1488\,MHz. The data
were calibrated for polarisation with the common single-axis model
\citep[e.g.][]{ovhb04}, with reference to a noise diode positioned
at 45 degrees to the linear feed probes. The \textsc{psrzap} and
\textsc{pazi} programs (\textsc{psrchive}'s RFI zapper) were used to
clean any RFI. We formed a two-dimensional analytic template by
fitting Gaussian components to the integration \citep[e.g.][]{kra94}
on MJD~56592 when the source was the brightest among our selected
epochs. With the remaining data we generated integrations of
$18-25$\,min length, with an ephemeris determined from the data
produced by the legacy BON backend with a baseline of 8\,yr. Most of
the preprocessing was conducted with the \textsc{psrchive} software
package \citep{hvm04}.

\subsection{Results} \label{ssec:result}
\begin{figure}
\centering
\includegraphics[scale=0.5,angle=-90]{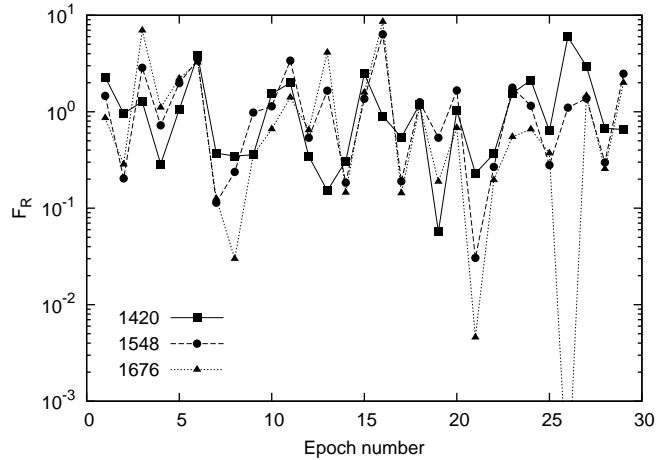}
\caption{Relative observed flux densities of the top three 128\,MHz
sub-bands (centred at 1420, 1548 and 1676\,MHz, respectively) with
respect to the bottom one (centred at 1292\,MHz) for all epochs.
\label{fig:frel}}
\end{figure}

The scintillation has been significant within our observing
frequency band. In Fig.~\ref{fig:frel} we divide the whole bandwidth
into four 128\,MHz sub-bands, and show the observed flux densities
of the top three divided by that of the bottom one for all epochs.
The ratios are seen to vary greatly for different epochs, mostly
within the scale of $10^{-2}-10$.

For comparison, we applied both the traditional template-matching
approach and the channelised DFT method to measure the ToAs of the
PSR~J1909$-$3744 integrations. Then we used the \textsc{tempo2}
software package \citep{hem06} to calculate the timing residuals
based on the aforementioned ephemeris, without fitting for any
parameters. When generating ToAs with the traditional method, we
dedispersed the data with DM values both retained from online
coherent dedispersion and determined by fitting for a constant DM
with multi-frequency ToAs from the same dataset.

\begin{figure}
\centering
\includegraphics[scale=0.5,angle=-90]{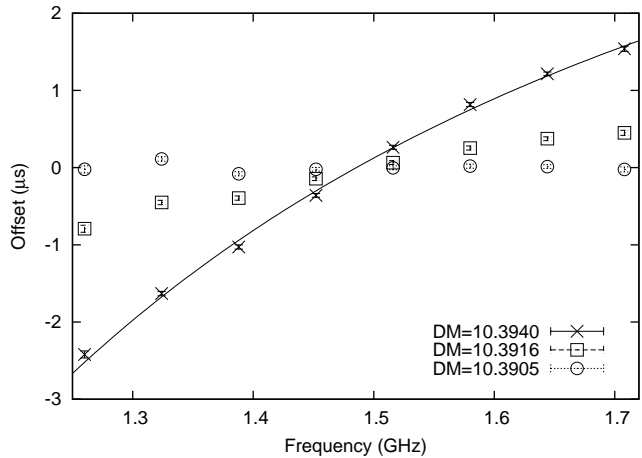}
\caption{Phase offsets of the observed profile on MJD~56592 at
different frequencies based on the DM values as in
Table~\ref{tab:res}. The solid line represents the best-fit
quadratic curve when using ${\rm DM}=10.3940$\,$\rm cm^{-3}pc$. The
value of $10.3905\,\rm cm^{-3}pc$ is the best-estimated DM with the
offsets. \label{fig:shtdm}}
\end{figure}

A brief summary of the timing results can be found in
Table~\ref{tab:res}. It can been seen that application of the
channelised DFT method (without fitting for $\Delta{\rm DM}$)
achieves the lowest weighted timing precision and improves the ToA
measurement uncertainties by $\sim20\%$ in median. Using the DM
measured from the dataset itself ($10.3916\,\rm cm^{-3}pc$) leads to
a similar timing precision, while retaining the DM used for coherent
dedispersion ($10.3940\,\rm cm^{-3}pc$) results in significantly
worse rms and more systematics. This can be explained by the fact
that a deviation of DM from the true value may induce additional
profile variation across the observing band and enhance the
scintillation effect. For demonstration, in Fig.~\ref{fig:shtdm} we
plot the phase offsets of the observed profile on MJD~56592 at
different frequencies, given the DM values as in
Table~\ref{tab:res}. Clearly, the DM value corresponding to lower
timing rms gives significantly less differential phase and thus
profile variation between frequencies. It is also interesting to
notice, that after subtracting all quadratic components the offsets
are mostly around zero. This indicates that the intrinsic profile
variation may not be significant within our observing bandwidth,
unless it actually appears as a quadratic drift against frequency.

\begin{table*}
\centering \caption{Statistical results of ToAs obtained by the
traditional template-matching approach (1D) with different DM values
for dedispersion, and the channelised DFT method without (2D$_{0}$)
and with $\Delta{\rm DM}$ fit (2D$_{1}$). The DM of 10.3916\,$\rm
cm^{-3}pc$ was achieved by fitting for a constant DM with
multi-frequency ToAs from the selected NUPPI dataset.}
\label{tab:res}
\begin{tabular}[c]{ccccc}
\hline
MJD                            &1D$_{0}$ &1D$_{1}$ &2D$_{0}$ &2D$_{1}$ \\
\hline
DM ($\rm cm^{-3}pc$)           &10.3940  &10.3916  &10.3940  &10.3940 \\
Max. $\sigma_{\rm ToA}$ (ns)   &934      &912      &839      &4094  \\
Min. $\sigma_{\rm ToA}$ (ns)   &39       &39       &30       &220   \\
Median $\sigma_{\rm ToA}$ (ns) &149      &149      &114      &1014  \\
Weighted rms (ns)              &610      &275      &247      &786   \\
Reduced $\chi^{2}$             &40.3     &8.13     &10.8     &1.85  \\
\hline
\end{tabular}
\end{table*}

\begin{figure}
\centering
\includegraphics[scale=0.48,angle=-90]{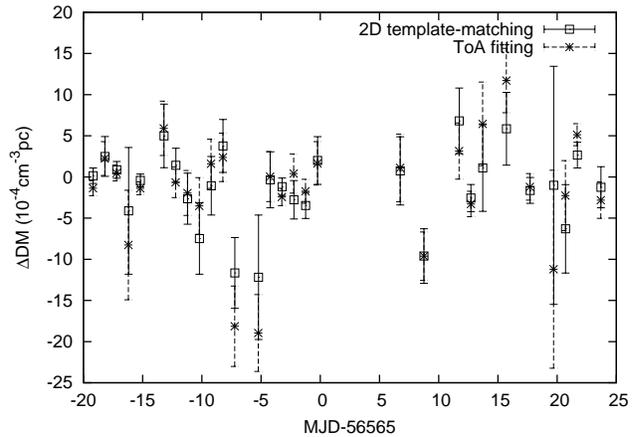}
\caption{Measured $\Delta{\rm DM}$ on each epoch with the
two-dimensional template-matching technique and the normal method
using multi-frequency ToAs. The measurements from template-matching
have rms and uncertainty median of $4.8\times10^{-4}$\,$\rm
cm^{-3}pc$ and $3.2\times10^{-4}$\,$\rm cm^{-3}pc$, respectively,
while those obtained based on multi-frequency ToAs achieve
$6.6\times10^{-4}$\,$\rm cm^{-3}pc$ and $2.9\times10^{-4}$\,$\rm
cm^{-3}pc$. \label{fig:1909dm}}
\end{figure}

Applying the channelised DFT method with $\Delta{\rm DM}$ fit
results in worse rms residuals, which is expected due to the high
correlation between $\Delta{\rm DM}$ and $\Delta\tau$ in the data as
indicated in Fig.~\ref{fig:f_bw}. Nevertheless, in this case
including $\Delta{\rm DM}$ into the template-matching fit may not be
necessary as DM variations are not significant within the observing
time. In Fig.~\ref{fig:1909dm} we plot the $\Delta{\rm DM}$
measurements at each epoch, achieved by both the channelised DFT
method and ToAs from multiple frequencies as in e.g. \cite{kcs+13}.
Here we divided the whole bandwidth into eight 64-MHz sub-bands to
create multi-frequency ToAs. It can be seen that the two methods
lead to consistent measurements, both in values and uncertainties.
Within the observing time there is also no evidence of gradual DM
variations above the detection threshold.

\section{Conclusion} \label{sec:conclu}
In this paper, we have described the channelised DFT method that can
conduct ToA measurements based on timing data with frequency
information. The functionality of the channelised DFT method has
been tested, and the potential for improvements to timing precision
has been demonstrated using simulated data. Furthermore, we have
applied the method to timing data of PSR~J1909$-$3744 with 512\,MHz
bandwidth centred at 1.45\,GHz, which removes systematics due to
scintillation effects enhanced by a faulty DM value and improves ToA
measurement uncertainties by $\sim20\%$ in median. Our approach has
also been shown to achieve measurements of $\Delta{\rm DM}$ on
epochs consistent with methods based on multi-frequency ToAs
\cite[as in e.g.][]{kcs+13}.

It is known that the observed dispersion delay can deviate from the
expected $\propto f^{-2}$ law coming from the cold plasma assumption
\citep[e.g.][]{ars95,kcs+13}. This phenomenon can potentially be
taken into account with the channelised DFT method, by simply
including a theoretical model in the merit function, i.e.,
Eq.~(\ref{eq:chisqr_DM}), and enabling more parameter fittings. For
this purpose, broadband observations would be highly requested to
break the degeneracy between fitted parameters and to achieve enough
sensitivity to measure the deviation.

Note that application of the channelised DFT method might not
achieve optimal modelling of DM variations as the measurements use
information only from a single observing session without any
interpolations over epochs like in other work
\citep{ktr94,kcs+13,lbj+14,lah+14}. Nevertheless, if the variation
timescale is well above the intervals between observing sessions,
the method can potentially be extended to combine data from
neighbouring epochs and perform a global fit for a unique offset in
DM, so as to increase the accuracy of determination. Besides, when
processing broadband timing data, the current interpolation
approaches do not fully use the information that data from different
frequencies are from a simultaneous observation. Therefore, it may
also be worth attempting to combine these two types of method, to
simultaneously obtain timing residuals and DM modelling directly
from timing data.

\section*{Acknowledgements}
We thank D.~Stinebring for help with simulations of the dynamic
spectra. We are also grateful to the anonymous referee who provided
constructive suggestions to significantly improve the paper. K.~Liu
is supported by the ERC Advanced Grant ``LEAP", Grant Agreement
Number 227947 (PI M.~Kramer). The Nan\c{c}ay Radio Observatory is
operated by the Paris Observatory, associated with the French Centre
National de la Recherche Scientifique.

\bibliographystyle{mnras}
\bibliography{journals_apj,psrrefs,modrefs,crossrefs}

\appendix
\section{Derivative calculations for L-M routine} \label{sec:app}
The L-M data modelling routine requires the input of a gradient of
$\chi^2$ and the curvature matrix. The calculation was done in
\cite{ptvf92} for a model of real numbers in a one-dimensional
array, and here we derive the expressions concerning complex numbers
and two dimensions. With the measurement $y_{j,k}=m_{j,k}+n_{j,k}i$,
and the model
$y(x_{j,k},\vec{a})=m(x_{j,k},\vec{a})+n(x_{j,k},\vec{a})i$, we have
\begin{eqnarray}
\chi^2(\vec{a})&=&\sum_{j,k}\left|\frac{m(x_{j,k},\vec{a})+n(x_{j,k},\vec{a})i-m_{j,k}-n_{j,k}i}{\sigma^2_{j,k}}\right|^2
\nonumber
\\
&=&\sum_{j,k}\frac{1}{\sigma^2_{j,k}}[m^2(x_{j,k},\vec{a})+m^2_{j,k}-2m(x_{j,k},\vec{a})m_{j,k} \nonumber \\
&&+n^2(x_{j,k},\vec{a})+n^2_{j,k}-2n(x_{j,k},\vec{a})n_{j,k}],
\end{eqnarray}
where $\vec{a}$ is an array containing the fitted parameters and
$i=\sqrt{-1}$. Hence, the gradient of $\chi^2$ is expressed by
\begin{eqnarray}
\frac{\partial\chi^2}{\partial\alpha_p}&=&\sum_{j,k}\frac{2}{\sigma^2_{j,k}}\left[\left(m(x_{j,k},\vec{a})
-m_{j,k}\right)\frac{\partial
m(x_{j,k},\vec{a})}{\partial\alpha_p}\right. \nonumber \\
&&\left. +\left(n(x_{j,k},\vec{a})-n_{j,k}\right)\frac{\partial
n(x_{j,k},\vec{a})}{\partial\alpha_p}\right],
\end{eqnarray}
and the second derivative matrix (Hessian matrix) is written as
\begin{eqnarray}
\frac{\partial\chi^2}{\partial\alpha_p\partial\alpha_q}&=&\sum_{j,k}\frac{2}{\sigma^2_{j,k}}[m(x_{j,k},\vec{a})\frac{\partial
m^2(x_{j,k},\vec{a})}{\partial\alpha_p\partial\alpha_q}+n(x_{j,k},\vec{a})\frac{\partial
n^2(x_{j,k},\vec{a})}{\partial\alpha_p\partial\alpha_q} \nonumber \\
&&+\frac{\partial
m(x_{j,k},\vec{a})}{\partial\alpha_p}\frac{\partial
m(x_{j,k},\vec{a})}{\partial\alpha_q}+\frac{\partial
n(x_{j,k},\vec{a})}{\partial\alpha_p}\frac{\partial
n(x_{j,k},\vec{a})}{\partial\alpha_q} \nonumber \\
&&-m_{j,k}\frac{\partial
m^2(x_{j,k},\vec{a})}{\partial\alpha_p\partial\alpha_q}-n_{j,k}\frac{\partial
n^2(x_{j,k},\vec{a})}{\partial\alpha_p\partial\alpha_q}].
\end{eqnarray}
The actual fitting routine uses the curvature matrix defined as
$\displaystyle\alpha_{pq}\equiv\frac{1}{2}\frac{\partial\chi^2}{\partial\alpha_p\partial\alpha_q}$.
After eliminating the second derivative terms so as to stabilise the
iterations \citep[explained in Chapter~15.5, ][]{ptvf92}, we then
have
\begin{equation}
\alpha_{pq}=\sum_{j,k}\frac{1}{\sigma_{j,k}^2}\left[\frac{\partial
m(x_{j,k},\vec{a})}{\partial\alpha_p}\frac{\partial
m(x_{j,k},\vec{a})}{\partial\alpha_q}+\frac{\partial
n(x_{j,k},\vec{a})}{\partial\alpha_p}\frac{\partial
n(x_{j,k},\vec{a})}{\partial\alpha_q}\right].
\end{equation}
Note that in our model
\begin{eqnarray}
\vec{a}&=&(b_1,...,b_{N_{\rm b}},\Delta\tau,\Delta{\rm DM}), \\
m_{j,k}&=&P_{j,k}\cos\theta_{j,k}, \\
n_{j,k}&=&P_{j,k}\sin\theta_{j,k}, \\
m(x_{j,k},\vec{a})&=&b_jS_{j,k}\cos(\phi_{j,k}+\tau_j), \\
n(x_{j,k},\vec{a})&=&b_jS_{j,k}\sin(\phi_{j,k}+\tau_j),
\end{eqnarray}
then $\chi^2$ becomes
\begin{equation}
    \chi^2=\sum_{j,k} \frac{P_{j,k}^2+b_j^2S_{j,k}^2 -2 b_j P_{j,k}S_{j,k}
    \cos\left(\phi_{j, k}-\theta_{j,k}+k \tau_j\right)} {\sigma_{j,k}^2}\,,
\end{equation}
and the gradient of $\chi^2$ is written as
\begin{equation}
\frac{\partial\chi^2}{\partial\alpha_p}=\left\{\begin{array}{cc}
\displaystyle\sum_i\frac{2b_pS^2_{p,k}-2P_{p,k}S_{p,k}\cos(\phi_{p,k}-\theta_{p,k}+k\tau_p)}{\sigma^2_{p,k}},
&1\leq p\leq N_{\rm b}\\
\displaystyle\sum_{j,k}\frac{2kb_jP_{j,k}S_{j,k}\sin(\phi_{j,k}-\theta_{j,k}+k\tau_j)}{\sigma^2_{j,k}}, &p=N_{\rm b}+1\\
\displaystyle\sum_{j,k}\frac{2k{D}_0b_jP_{j,k}S_{j,k}\sin(\phi_{j,k}-\theta_{j,k}+k\tau_j)}{\sigma^2_{j,k}f^2_j},
&p=N_{\rm b}+2
\end{array}\right.
\end{equation}
and the curvature matrix is
\begin{equation}
\alpha_{pq}=\left\{\begin{array}{cc}
   \displaystyle \sum_k\frac{S^2_{p,k}}{\sigma^2_l}, &1\leq p,q\leq N_{\rm b},~p=q  \\
   \displaystyle 0, &1\leq p,q\leq N_{\rm b},~p\neq q \\
   \displaystyle 0, &1\leq p\leq N_{\rm b},~N_{\rm b}\leq q\leq N_{\rm
   b}+2 \\
   \displaystyle \sum_{j,k}\frac{k^2b^2_jS^2_{j,k}}{\sigma^2_j}, &p=q=N_{\rm b}+1\\
   \displaystyle
   \sum_{j,k}\frac{k^2D_0b^2_jS^2_{j,k}}{\sigma^2_jf^2_j}, &p=N_{\rm
   b}+1,~q=N_{\rm b}+2 \\
   \displaystyle\sum_{j,k}\left(\frac{kD_0b_jS_{j,k}}{\sigma_jf_j}\right)^2, &p=q=N_{\rm b}+2.
  \end{array}\right.
\end{equation}
\end{document}